# Uncertainty of Convection

L.S. Yao and S. Ghosh Moulic

Department of Mechanical and Aerospace Engineering
Arizona State University
Tempe, AZ  85287-6106

## ABSTRACT

The nonlinear development of finite amplitude disturbances in mixed convection flow in a heated vertical annulus is studied by direct numerical simulation. The unsteady Navier Stokes equations are solved numerically by a spectral method for different initial conditions. The results indicate that the equilibrium state of the flow is not unique, but depends on the amplitude and wavenumber of the initial disturbance. In all cases, the equilibrium state consists of a single dominant mode with the wavenumber $k_f$, and its superharmonics. The range of equilibrium wavenumbers $k_f$ was found to be narrower than the span of the neutral curve from linear theory. Flows with wavenumbers outside this range, but within the unstable region of linear theory are found to be unstable and to decay, but to excite another wave inside the narrow band. This result is in agreement with the Eckhaus and Benjamin-Feir sideband instability. The results also show that linearly stable long and short waves can also excite a wave inside this narrow band through nonlinear wave interaction. The results suggest that the selection of the equilibrium wavenumber $k_f$ is due to a nonlinear energy transfer process which is sensitive to initial conditions.

The consequence of the existence of nonunique equilibrium states is that the Nusselt number cannot always be expressed uniquely as a function of appropriate dimensionless parameters such as the Reynolds, Prandtl and Rayleigh numbers. Any physical quantity transported by the fluid, such as heat and salt, can at best be determined within the limit of *uncertainty* associated with nonuniqueness. This uncertainty should be taken into account when using any accurately measured values of the heat transfer rate since it is only one of the many possible states for the controllable conditions and geometry. Extrapolating this fact to turbulence, it is our opinion, since the *time* average will depend on the initial condition, it will not equal to the *ensemble* average even for stationary turbulence. This is because that the mean flow is not unique for a given

Reynolds or Rayleigh number.  Consequently, the ergodic hypothesis is not valid.  From an application point of view, only time average has physical significance.

**NOMENCLATURE**

| | |
|---|---|
| $A_n$ | initial disturbance amplitude of n-th Fourier mode |
| C | constant determining average axial velocity of isothermal flow through annulus |
| d | distance between cylinders |
| E | disturbance kinetic energy |
| E(k) | kinetic energy of Fourier mode with wavenumber k |
| $\mathbf{e}_z$ | unit vector in z direction |
| g | gravitational acceleration |
| G | axial pressure gradient |
| h | local heat transfer coefficient |
| k | axial wavenumber |
| $k_f$ | final equilibrium wavenumber |
| $\bar{k}$ | thermal conductivity of fluid |
| $N_y$ | number of collocation points in the radial direction |
| $N_z$ | number of collocation points in the axial direction |
| $N_\varphi$ | number of collocation points in the azimuthal direction |
| Nu | Nusselt number |
| $Nu_z$ | local Nusselt number |
| p | pressure |
| Pr | Prandtl number |
| $q_w$ | local heat flux at inner wall |
| Ra | Rayleigh number |
| Re | Reynolds number |
| r | radial coordinate |
| $r_1$ | radius of inner cylinder (dimensional) |
| $r_2$ | radius of outer cylinder (dimensional) |
| $r_i$ | dimensionless radius of inner cylinder |
| $r_o$ | dimensionless radius of outer cylinder |
| t | time |
| T | dimensional temperature |
| $T_b$ | bulk temperature of fluid |
| $T_k(y)$ | k-th degree Chebyshev polynomial |
| $T_o$ | upstream reference temperature |
| $T_w$ | temperature of inner wall |
| u | radial velocity |
| $\hat{u}$ | discrete Fourier transform of radial velocity |
| $\hat{u}_n$ | radial velocity predicted by linear stability theory |
| **u** | =(u,v,w), velocity vector |
| v | azimuthal velocity |
| $\hat{v}_n$ | azimuthal velocity predicted by linear stability theory |
| w | axial velocity |
| $\hat{w}_n$ | axial velocity predicted by linear stability theory |
| W | average axial velocity for isothermal flow through an annulus |
| $W_0$ | fully developed axial velocity profile |
| y | normalized radial coordinate |
| z | axial coordinate |

Greek symbols

| | |
|---|---|
| α | fundamental wavenumber |
| β | coefficient of thermal expansion |
| φ | azimuthal angle |
| η | radius ratio of annulus |
| κ | thermal diffusivity of fluid |
| λ | wavelength of computational box |
| μ | axial temperature gradient |
| ν | kinematic viscosity of fluid |
| θ | dimensionless temperature |
| $\theta_0$ | fully developed temperature distribution |
| ρ | density of fluid |
| π | $= p + \frac{1}{2}|\mathbf{u}|^2$ |
| ϖ | $= \nabla \times \mathbf{u}$ |

1.  **INTRODUCTION**

The study of nonisothermal flows in heated ducts is of fundamental importance in engineering applications, and has been the subject of many experimental and theoretical investigations. The final outcome of most of these investigations is a prediction of the Nusselt number for different geometries and heating conditions. The Buckingham pi Theorem implies that the Nusselt number may be expressed deterministically as a function of appropriate dimensionless control parameters such as the Reynolds, Rayleigh and Prandtl numbers. Thus, the results of experiments conducted using geometrically similar small-scale prototype models can be used to predict the heat-transfer rates in real engineering situations. This paper demonstrates that the Nusselt number cannot always be determined uniquely even if the control parameters are fixed uniquely and the flow has reached an equilibrium state, due to the possible existence of multiple equilibrium states. The implication of the existence of nonunique equilibrium states is that any physical quantity transported by the fluid, such as heat and salt, can at best be determined within the limit of uncertainty associated with nonuniqueness. This uncertainty can occur even in the laminar flow regime after its first bifurcation point.

Mixed convection flow in a heated vertical annulus is used as a model problem to illustrate this uncertainty principle. It is common practice to treat such flows as parallel flows. The parallel flow assumption simplifies the analysis considerably, and the velocity and temperature fields can be easily predicted as functions of the transverse coordinate. However, recent work has shown that fully developed mixed convection flows in vertical ducts are highly unstable due to thermally-induced instabilities [1-4]. The presence of instabilities in nonisothermal flows in heated vertical pipes has been observed experimentally by Hanratty, Rosen & Kabel [5], Kemeny & Somers [6] and Scheele & Hanratty [7]. These instabilities lead to significant increases in the heat transfer rates above those predicted by parallel flow models. When the flow is stably stratified, with the buoyancy forces aiding the fluid motion, they observed that the initial transition resulted in a new equilibrium non-parallel flow. However, when the flow is unstably stratified, with the buoyancy forces opposing the fluid motion, the transition to turbulence can be abrupt. Thus, the bifurcation is supercritical when buoyancy forces aid the fluid motion, and may be subcritical when the buoyancy forces oppose the fluid motion. Similar instabilities have been observed by Maitra & Subba Raju [8] for flow in a heated vertical annulus. They also found that the instabilities resulted in heat transfer rates which were much higher than those predicted by a parallel-flow model. The evolution of finite-amplitude disturbances in this flow situation was studied theoretically by using weakly nonlinear instability theory [9]. The results indicate that the bifurcation is supercritical when buoyancy forces aid the fluid motion and are in agreement with the experiments of Maitra & Subba Raju.

In this investigation, the nonlinear development of disturbances in mixed convection flow in a heated vertical annulus is simulated numerically by solving the Navier-Stokes equations using a spectral method. The time-dependent Navier-Stokes system is solved starting with different initial conditions. The results show that the equilibrium state depends on the amplitude and shape of the initial disturbance, and is not unique. In all cases, the final equilibrium state is a monochromatic traveling wave consisting of a single dominant mode with a wavenumber $k_f$, together with its harmonics, and an induced mean flow distortion. The range of equilibrium wavenumbers $k_f$ was found to be a subset of the linearly unstable band of wavenumbers. Flows with wavenumbers outside this subset but within the linearly unstable band of wavenumbers were found to be unstable and to decay, the energy being transferred to a mode with wavenumber inside this subset, which is excited through nonlinear interaction. This result is in agreement with the Eckhaus and Benjamin-Feir sideband instability [10]. Numerical simulations starting with initial conditions consisting of a single dominant mode with wavenumbers lying in the linearly stable range also resulted in an equilibrium state with a final dominant wavenumber $k_f$ inside this subset. Similar results were obtained for Taylor-Couette flow by Yao & Ghosh Moulic [11], using a weakly nonlinear analysis with a continuous spectrum. They represented the disturbance by a Fourier integral, and derived an integro-differential Landau equation for the amplitude density of the wave-components of a continuous spectrum of waves. Numerical integration of this Landau equation showed that the final equilibrium state was not unique, but depended on the waveform of the initial disturbance and the initial wave amplitudes, as observed experimentally [12-14]. These results suggest that the selection of the equilibrium wave-number is due to a nonlinear energy transfer process, which is sensitive to initial conditions.

The consequence of the existence of nonunique equilibrium states is that the Nusselt number cannot always be uniquely determined as a function of appropriate dimensionless control parameters such as the Reynolds, Rayleigh and Prandtl numbers, as implied by the Buckingham pi theorem. Any physical quantity such as heat, species etc. transported by the fluid can at best be determined within the limit of uncertainty associated with nonuniqueness. This uncertainty can not be ignored in practice.

The numerical results obtained in this investigation indicate that nonlinear effects induce a large distortion of the mean flow. The kinetic energy of the distorted mean flow is found to be the dominant component of the disturbance kinetic energy, and much higher than the kinetic energy of the fundamental dominant wave in the equilibrium state. This implies that classical weakly nonlinear theories of monochromatic waves [15,16] as well as slowly-varying wave packets [17-19] which assume *a priori* that the mean flow distortion is induced by the fundamental wave and is a much smaller order effect, is not valid. This is because the classical theories are for monochromatic waves, and do not consider the nonlinear energy transfer among different waves. This drawback can be overcome by reformulating the problem with a continuous spectrum [20]. The analysis in [20] is an extension of the weakly nonlinear theory of a continuous spectrum of stationary waves [11] to traveling waves. The disturbance is represented by a Fourier integral over all possible wavenumbers. The Fourier components are expanded in a series of the linear stability eigenfunctions. The eigenfunction expansion reduces the Navier-Stokes equations to a set of nonlinearly coupled integrodifferential equations for the amplitude density function of a continuous spectrum without any approximation. The equations describing the evolution of monochromatic waves and slowly varying wavepackets of classical weakly nonlinear theories are special limiting cases of the integrodifferential equations in a parameter range close to the onset of linear instability with a single eigenmode for the waves. Comparison with the numerical solution of the Navier-Stokes equations, however, indicates that the range of validity of the classical weakly nonlinear theories is very small. There are cases where the classical theories predict qualitatively incorrect results. The solution of the integrodifferential equations with 20 eigenmodes, on the other hand, agrees with the results of the direct numerical simulation of the Navier-Stokes equations presented in this paper. This is because the integrodifferential equations are equivalent to the Navier-Stokes equations. Thus, the solution of the integrodifferential equations is an exact solution of the Navier-Stokes equations. The CPU time required for the solution of the integrodifferential equations is only one-fourth of the CPU time required for the direct simulation of the Navier-Stokes equations using a Fourier-Chebyshev spectral method in this paper, but requires more computer memory.

2.  ANALYSIS

We consider the flow in a heated vertical annulus, driven by an externally applied pressure gradient, as illustrated in Figure 1. A constant vertical temperature gradient is maintained at the inner cylinder, and the outer cylinder is insulated. Let $(r, \phi, z)$ denote cylindrical polar coordinates, with the z-axis aligned with the common axis of the cylinders, and let $(u, v, w)$ denote the corresponding velocity components. The equations describing the flow are the continuity, Navier-Stokes and energy equations. Using the Boussinesq approximation, these equations may be written in dimensionless form as

$$\nabla \bullet \mathbf{u} = 0$$

$$\frac{\partial \mathbf{u}}{\partial t} = \frac{C}{\text{Re}}\mathbf{e}_z - \nabla \pi + \mathbf{u} \times \varpi + \frac{1}{\text{Re}}\nabla^2 \mathbf{u} - \frac{Ra}{\text{Re}}\theta \mathbf{e}_z \qquad (1)$$

$$\frac{\partial \theta}{\partial t} + \mathbf{u} \bullet \nabla \theta = \frac{1}{\text{Re Pr}}\nabla^2 \theta + \frac{w}{\text{Re Pr}},$$

where $\mathbf{u} = (u,v,w)$, $\varpi = \nabla \times \mathbf{u}$, $\pi = p + \frac{1}{2}|\mathbf{u}|^2$, and $\mathbf{e}_z$ is the unit vector along z direction. All lengths have been scaled by the distance between the cylinders, $d = r_2 - r_1$, where $r_1$ is the radius of the inner cylinder and $r_2$ is the radius of the outer cylinder. Our choice of the velocity scale is based on the applied pressure gradient, G. For isothermal flow through an annulus, the average axial velocity is $W = G\,d^2 / (C\rho\nu)$, where,

$$C = \frac{8(1-\eta)}{\frac{1+\eta^2}{1-\eta} + \frac{1+\eta}{\ln \eta}},$$

$\eta = r_1 / r_2$ is the radius ratio of the annulus, $\rho$ is the density of the fluid, and $\nu$ is the kinematic viscosity. This provides a natural velocity scale for this problem. A nondimensional pressure is defined by

$$p = \frac{\bar{p} + G\,d\,z}{\rho W^2},$$

where $\bar{p}$ is the (dimensional) pressure fluctuation. The time is scaled by d/W. The temperature of the inner cylinder increases linearly with the axial coordinate from an upstream reference temperature, $T_o$, as $T_w = T_o + \mu\,d\,z$, where $\mu$ is the constant vertical temperature gradient. In the limiting case of a fully developed flow, this simulates a uniform heat flux thermal boundary condition on the inner cylinder. A dimensionless temperature has been defined by

$$\theta = \frac{T_w - T}{\mu d \, \text{Re} \, \text{Pr}}$$

The parameters in this problem are the Reynolds number, Re = Wd/ν = Gd³ / (C ρ ν²), the Prandtl number, Pr = ν / κ , and the Rayleigh number Ra = μ β g d⁴ / (ν κ),  Here β is the coefficient of thermal expansion, κ is the thermal diffusivity, and g is the gravitational acceleration.  The boundary conditions are

$$u = v = w = \theta = 0 \text{ when } r = r_i,$$

and (2)

$$u = v = w = \frac{\partial \theta}{\partial r} = 0 \text{ when } r = r_0,$$

where $r_i = r_1 / d$ and $r_o = r_2 / d$.

The equations (1) admit a steady parallel-flow solution

$$u = v = 0, \, w = W_0(r), \, \theta = \theta_0(r),\qquad(3)$$

where the functions $W_0(r)$ and $\theta_0(r)$ are solutions of the following system of equations:

$$\frac{d^2 W_o}{dr^2} + \frac{1}{r}\frac{dW_o}{dr} - Ra\,\theta_o = -C,$$

(4)

$$\frac{d^2 \theta_o}{dr^2} + \frac{1}{r}\frac{d\theta_o}{dr} + W_o = O.$$

The stability of this parallel-flow may be studied by superposing a disturbance on this flow.  The finite-amplitude evolution of such disturbances has been studied by Yao & Rogers [9] using a perturbation method.  In this study, we solve the system of equation (1) as an initial value problem using a spectral method.  The spatial discretization is based on Fourier expansions in the axial and azimuthal directions and Chebyshev polynomials in the radial direction. The dependent variable are represented by expansions of the form

$$u(r,\phi,z,t) = \sum_{n=-\frac{N_z}{2}}^{\frac{N_z}{2}-1} \sum_{m=-\frac{N_\phi}{2}}^{\frac{N_\phi}{2}-1} \hat{u}(r,m,n,t) e^{i[m\phi+n\alpha z]},$$

$$\hat{u}(r,m,n,t) = \sum_{k=o}^{N_y} \tilde{u}(k,m,n,t) T_k(y),$$
(5)

where $\alpha$ is the fundamental wavenumber in the axial direction, $T_k(y)$ is the $k^{th}$ degree Chebyshev polynomial defined by $T_k(y) = \cos(k \cos^{-1} y)$, y is a normalized radial coordinate defined by

$$y = \frac{2(r-r_i)}{r_o - r_i} - 1,$$
(6)

and $N_y$, $N_z$ and $N_\phi$ are the number of collocation points in the radial, axial and azimuthal directions respectively. The collocation points are

$$y_j = \cos\left(\frac{\pi j}{N_y}\right), j = 0, 1, ..., N_y,$$

$$z_j = \frac{2\pi}{\alpha N_z}, j = 0, 1..., N_z - 1,$$

$$\phi_j = \frac{2\pi j}{N_\phi}, j = 0, 1, ..., N_\phi - 1.$$
(7)

This choice of collocation points yields 'spectral accuracy' and allows fast transformation between physical space and wave space. Time-differencing was done using a Crank-Nicholson scheme for the diffusion terms and second-order Adams-Bashforth for the convection and body force terms. The rotation form of the Navier-Stokes equations is preferred for the numerical simulations because, as noted by Orszag[21], the use of this form guarantees that Fourier collocation methods conserve kinetic energy and ensures that the nonlinear terms do not cause numerical instability. Time-differencing errors are reduced by using coordinate system moving with the (approximate) phase speed of the fundamental wave, although, because of Galilean invariance, this is equivalent to a calculation in a frame of reference which is at rest. The momentum equations are decoupled by solving a Poisson equation for the pressure. The correct boundary conditions for the pressure, consistent with a divergence-free velocity field at the solid boundaries, are obtained by an influence matrix technique [22,23]. The numerical procedure requires the solution of a sequence of Helmholtz equations at each time step. These equations are solved by a preconditioned minimum residual method [24].

## 3. RESULTS AND DISCUSSION

Results have been obtained for an annulus with a radius ratio $\eta = 0.375$, Re = 1000, Ra=200 and Pr = 6. The radius ratio of 0.375 was chosen so that the results could be compared with the experiments of Maitra & Subba Raju [8]. The critical Rayleigh number at the onset of instability for this flow configuration is $Ra_c = 89$ [9]. Linear stability analysis indicates that at Ra=200, Re = 1000 and Pr = 6, the parallel basic flow is unstable to disturbances with wavenumbers lying between 0.23 and 1.13. The time-dependent Navier Stokes system (1) was solved numerically for different initial conditions. The initial conditions used for the computations were of the form

$$u = \sum_{n=-\frac{N_z}{2}}^{\frac{N_z}{2}-1} A_n \hat{u}_n(r) e^{in\alpha z},$$

$$w = W_o(r) + \sum_{n=-\frac{N_z}{2}}^{\frac{N_z}{2}-1} A_n \hat{w}_n(r) e^{in\alpha z}, \tag{8}$$

$$\theta = \theta_o(r) + \sum_{n=-\frac{N_z}{2}}^{\frac{N_z}{2}-1} A_n \hat{\theta}_n(r) e^{in\alpha z},$$

where $W_o(r)$ and $\theta_0(r)$ are the fully-developed velocity and temperature profiles respectively, $\hat{u}_n, \hat{w}_n$ and $\hat{\theta}_n$ are the linear-instability eigenfunctions for an axial wavenumber $k = n\alpha$, and $A_n$ is the initial disturbance amplitude for the mode with wavenumber $k = n\alpha$. The computations were done on the CRAY C-90 supercomputer at the Pittsburgh Supercomputing Center. In order to minimize the computer time, most of the computations were done for the axisymmetric case, using $\alpha = 0.25$ and $N_z = 36$. Some of the computations have been done on a finer grid using $\alpha = 0.1$ and $N_z = 96$. The asymmetric azimuthal modes are linearly stable at the parameters used in this study, and although they can be excited through nonlinear interactions, we do not expect this to happen at $Ra = 200$. The validity of this assumption has been checked for one of the cases by solving the three-dimensional Navier-Stokes system, using 8 Fourier modes in the azimuthal direction. The asymmetric azimuthal modes were not excited in this computation, that is, the flow remained axisymmetric. Adequate spatial resolution was ensured by monitoring the kinetic energies of the highest Fourier modes. For axisymmetric flow, the kinetic energy of the $k^{th}$ Fourier mode is given by

$$E(k) = \int_{r_i}^{r_o} r \left[ |\hat{u}(r,0,k,t)|^2 + |\hat{w}(r,0,k,t)|^2 \right] dr, \qquad (9)$$

for $k \neq 0$, and

$$E(0) = \frac{1}{2} \int_{r_i}^{r_o} r \left[ |\hat{w}(r,0,0,t)|^2 - W_o^2(r) \right] dr, \qquad (10)$$

for $k = 0$. Equation (9) accounts for the energy in both modes $\pm k$. The kinetic energy of the fully-developed flow is subtracted from the mean-flow kinetic energy in (10) so that the total disturbance kinetic energy is given by

$$E = \sum_{k=0}^{\frac{N_z}{2}-1} E(k). \tag{11}$$

Aliasing errors resulting from severe truncation in the number of Fourier modes can result in an artificial curl in the high-wavenumber end of the energy spectrum. This was found to occur in our computations, and is illustrated in Figure 2 which shows the spectrum of kinetic energy at time t = 800, for a numerical simulation using $\alpha = 0.25$ in which the mode k = 0.75 was given an initial amplitude of 0.005 at time t = 0, and the other modes were given small initial amplitudes of $10^{-10}$. The spectrum for the aliased calculation curls upwards at the high wavenumber end. In order to get accurate results, we eliminate the aliasing errors by padding, using the two-third rule [24]. The results of the de-aliased calculation at t = 800, starting with the same initial conditions is also shown in Figure 2. The spectra in this case did not exhibit the artificial curl at the high wave-number end. It is worthwhile to note that in the de-aliased calculation, the energies in the modes k=0.25, 0.5, 1, 1.25 etc. which are not integral multiples of the fundamental mode k=0.75, was negligible. In the aliased calculation, on the other hand, all these modes were excited due to the aliasing error. This indicates that in this problem, aliasing errors can cause serious problems, and need to be eliminated.

Figure 3(a) shows the evolution of the kinetic energy of the dominant wave components for a numerical simulation with $\alpha = 0.25$ in which the mode k = 0.75 was given an initial amplitude of 0.005, and the other modes were given small initial amplitudes of $10^{-10}$. The mode k = 0.75 is linearly unstable, and grows by obtaining energy from the mean flow. Nonlinear interactions generate the harmonics k = 1.5, 2.25 and 3, and induce a mean flow distortion (k = 0). As the amplitude of the mode k = 0.75 increases, nonlinear effects become more important, and alter the linear growth rate, causing the mode to decay and eventually reach an equilibrium state. The equilibrium state is a monochromatic traveling wave in which the fundamental mode k = 0.75 remains the dominant mode, while its superharmonics, generated through nonlinear interaction, have smaller amplitudes. The kinetic energy of the third and fourth harmonics ( $k = 2.25$ and $k = 3$ ) are much smaller than the energies of the fundamental mode $k = 0.75$, and the second harmonic $k = 1.5$; these modes have not been plotted in Figure 3(a). It is worth noting that the kinetic energy associated with the mean flow distortion is higher than the kinetic energy of the fundamental mode $k = 0.75$. This implies that classical weakly nonlinear theories which assume *a priori* that the order of magnitude of the mean flow distortion is smaller than that of the fundamental wave, is not valid for this problem. In order to check the accuracy of the numerical result, the calculation was repeated using a finer grid with $\alpha = 0.05$, and 8 Fourier modes in the azimuthal direction. This computation required a lot of computer time, and was carried out only to a final time $t = 500$. The result of the computation using the fine grid is superimposed on Figure 3(a). The difference in the two computations can hardly be noticed in the scale of Figure 3(a). This indicates that the coarse grid has adequate resolution. It is worth pointing out that the energies in the Fourier modes which were not integral multiples of 0.75 were negligible, that is these modes were not excited. Hence, a computation which does not include these modes produces the same result as a computation which includes all these modes.

Figure 3(b) shows the results of a numerical simulation with $\alpha = 0.25$ in which the mode $k = 0.5$ was given an initial amplitude of $10^{-2}$ and the other modes were given initial amplitudes of $10^{-10}$. The equilibrium state in this case is a monochromatic wave with a dominant mode $k = 0.5$, and its small-amplitude super-harmonics.

Figure 3(c) shows the results of a numerical simulation with $\alpha = 0.25$, starting with a single dominant mode with wavenumber $k = 0.25$ at time $t = 0$. The mode $k = 0.25$ is linearly unstable, and the weakly nonlinear theory of monochromatic waves [15,16] predicts a supercritical equilibrium state for this mode. However, as indicated in Figure 3(c), the mode $k = 0.25$ grows initially and then decays to zero, while its harmonic $k = 0.5$, excited through nonlinear interaction, grows and reaches a finite-amplitude supercritical equilibrium state. This indicates that the equilibrium state predicted by the weakly nonlinear theory of monochromatic waves for the mode k = 0.25 is unstable. This result is in agreement with the Eckhaus and Benjamin-Feir side-band instability.

The results of a numerical simulation with $\alpha$ = 0.25, is which the mode k = 1 was given an initial amplitude of 0.005, and the other modes were given small initial amplitudes of $10^{-6}$, is shown in Figure 3(d). The mode k = 1 is linearly unstable and grows initially according to linear theory, as nonlinear effects are initially small. As its amplitude increases, nonlinear effects become important, and cause it to decay. The subharmonic mode k = 0.5, excited by nonlinear interaction grows and reaches an equilibrium amplitude which is higher than that of the mode k=1. The mode k = 1 remains in the final equilibrium state, but is not the dominant mode. This result agrees with the Eckhaus and Benjamin-Feir sideband instability. It is worth noting that subharmonic transitions such as this may play an important role in the transition to turbulence at higher Rayleigh numbers.

Figure 3(e) shows the results of a numerical simulation with $\alpha$ = 0.1, in which the mode k = 0.2 was given an initial amplitude of 0.005, and the other modes were given small initial amplitudes of $10^{-6}$. The mode k = 0.2 is linearly stable and decays to zero, as it transfers energy to the mean flow. However, nonlinear effects excite the mode k = 0.4, which grows and finally becomes the dominant wave in the equilibrium state. This result suggests that nonlinear wave interaction may be a credible wavelength adjustment mechanism for boundary-layer receptivity, which may serve to explain how a short Tollmien Schlichting wave in a boundary layer is excited by long-wavelength free-stream noise [25].

Figure 3(f) shows the results of a numerical simulation with $\alpha = 0.25$, starting with a single dominant mode k = 1.25, which was given an initial amplitude of 0.005. The mode k = 1.25 is linearly stable and decays to zero. However, nonlinear interactions transfer energy to subharmonic modes, exciting the mode k = 0.5, which grows and reaches an equilibrium state. The results shown in Figures 3(e) and 3(f) indicate that linearly stable long and short waves can play an important role in the transition process from laminar to turbulent flow, although they may eventually decay to zero, by transferring energy to other modes through nonlinear wave-interaction.

Figure 3(g) shows the dominant waves which evolve in a numerical simulation with $\alpha=0.25$, starting with a uniform broad-band spectrum with initial amplitude 0.001 at time t=0. The dominant wave in the equilibrium state in this case is the mode k = 0.5, which is the wavenumber closest to the minimum critical wavenumber at the onset of instability. Numerical simulations starting with a uniform initial amplitude of $10^{-6}$ also resulted in the same equilibrium state.

The selection of the equilibrium wavenumber is due to nonlinear wave interactions. The numerical results suggest the selection principles are similar to Taylor-Couette flows[11]:

1) When the initial disturbance consists of a single dominant wave within the unstable region, the initial wave remains dominant in the final equilibrium state. Consequently, for a slowly accelerating cylinder, the critical wave is likely to be dominant.

2) When the initial condition consists of two waves with finite amplitudes in the unstable region, the final dominant wave is the one with the higher initial amplitude. If the two waves have the same initial finite amplitude, the dominant wave seems to be the one closer to the critical wave. On the other hand, if the initial amplitudes are very small, the faster growing wave becomes dominant.

3) When the initial disturbance is a uniform broad-band spectrum, the final dominant wave is the fastest linearly growing wave, if the initial amplitude is small. On the other hand, if the uniform noise level is not small, the critical wave is the dominant equilibrium one.

An average Nusselt-number has been calculated, defined by, $Nu = \frac{1}{\lambda} \int_0^\lambda Nu_z \, dz$, where $\lambda = \frac{2\pi}{\alpha}$ is the wavelength of the computational box, $Nu_z = \frac{hd}{\bar{k}}$ is the local Nusselt number, $\bar{k}$ is the thermal conductivity of the fluid and h is the local heat-transfer coefficient, given by $h = \frac{q_w}{T_w - T_b}$ Here, $q_w$ and $T_w$ are the local heat-flux and temperature of the inner wall respectively, and $T_b$ is the bulk temperature of the fluid. The equilibrium values of the Nusselt number is shown in Table 1 for different equilibrium states. The increase in the Nusselt number due to the flow bifurcation, relative to its basic state value, is also shown in Table 1. As the table indicates, the increase in the Nusselt number varies from 8.5% for an equilibrium wavenumber of 0.4 to 12.1% for an equilibrium wavenumber of 0.75. The Nusselt number is plotted in Figure 4 as a function of the Rayleigh number. The results obtained by Yao & Rogers [9], using the weakly nonlinear theory of monochromatic waves is shown in Figure 4, and compared with the experimental data of Maitra & Subba Raju [8]. The variation in the Nusselt number for different equilibrium states predicted by the direct numerical simulations in the current investigation is indicated by the vertical line at Ra = 200. The numerical value of the uncertainty in the Nusselt number is small for the present case. However, the underlying principle has far reaching consequences. It is worth pointing out that in the current problem, the linearly unstable range of wavenumbers at Ra = 200 is small. As Ra increases, the linearly unstable range will increase accordingly. Also, many problems where the linearly unstable span of the neutral curve is large near the first bifurcation point, such as mixed convection flows in a heated vertical pipe [26, 27], the uncertainty in the Nusselt number is likely to be much larger and may not be ignored in practice. Since the same phenomenon has been found for the Taylor-Couette flow [11] and the mixed convection in this paper, we believe that the uncertainty phenomenon is generic and is applicable to all fluid flows, isothermal or non-isothermal, after their first bifurcation points.

## 4. CONCLUSIONS

The current investigation demonstrates that the supercritical equilibrium state of traveling waves bifurcating from fully-developed mixed-convection flow in a heated vertical annulus is not unique. Numerical simulations of the time-dependent Navier-Stokes equations with different initial conditions show that the final equilibrium state of the flow depends on the waveform of the initial disturbance. The results suggest that as the flow evolves from a given initial state, it selects an equilibrium wavenumber from a possible range of allowed wavenumbers. The selection of the equilibrium wavenumber is governed by a nonlinear energy transfer process which is sensitive to initial conditions. The allowed range of equilibrium wavenumbers is a subset of the linearly unstable span of wavenumbers. An important implication of the existence of non-unique equilibrium states is that the Nusselt number can be determined only within the limit of uncertainty associated with nonuniqueness.

It is suitable to speculate what the averaging process would be for turbulence since the convection has multiple solutions in a continuous range of spectra after the first bifurcation point. Since the *time* average mean flow and associated turbulent quantities depend on the initial conditions and are not unique, their values are not equal to the *ensemble* averages even for stationary turbulence. From an application point of view, only time average has physical significance. The ensemble average would be the averaged value of the all possible time averages which can be measured independently.

The results of the paper were computed by Cray C90 at the Pittsburgh Supercomputing Center. Their quick approval of the grant, CTS930015P, and their efficient support make the computation possible and is gratefully acknowledged. We also want to thank Mr. Bruce Tachoir of Supercomputer Consulting Center at Arizona State University for his efforts in helping us to become familiar with the Cray system and in optimizing our computer code within minimum possible period of time.

Table 1. Variation of Nusselt number with wavenumber.

| $k_f$ | 0.4 | 0.5 | 0.75 |
|---|---|---|---|
| Nu | 4.66 | 4.74 | 4.8 |
| Percentage of increase | 8.5 | 10.4 | 12.1 |

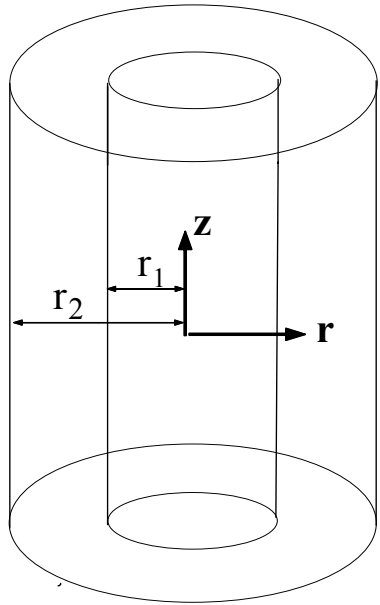

**Figure 1. Geometry and coordinates**

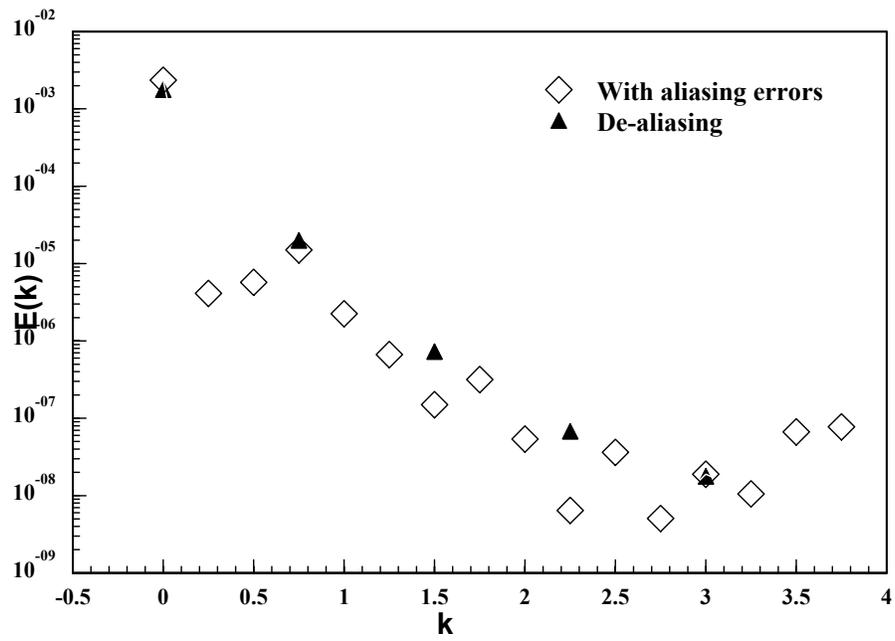

**Figure 2. Kinetic energy spectrum, illustrating aliasing errors**

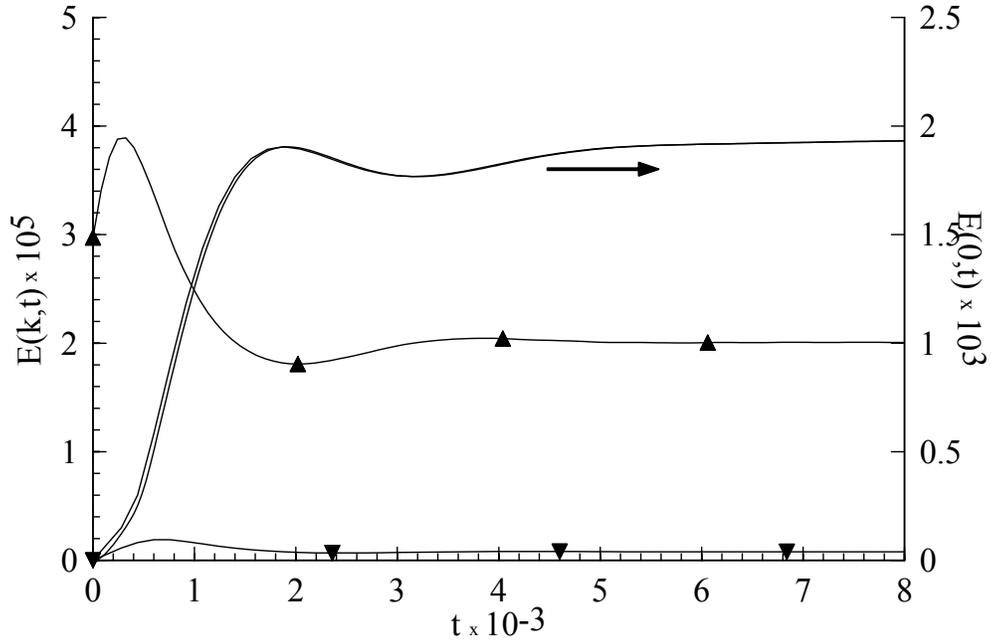

Figure 3a. Evolution of the kinetic energy spectrum for a single initial mode with wavelength k=0.75.

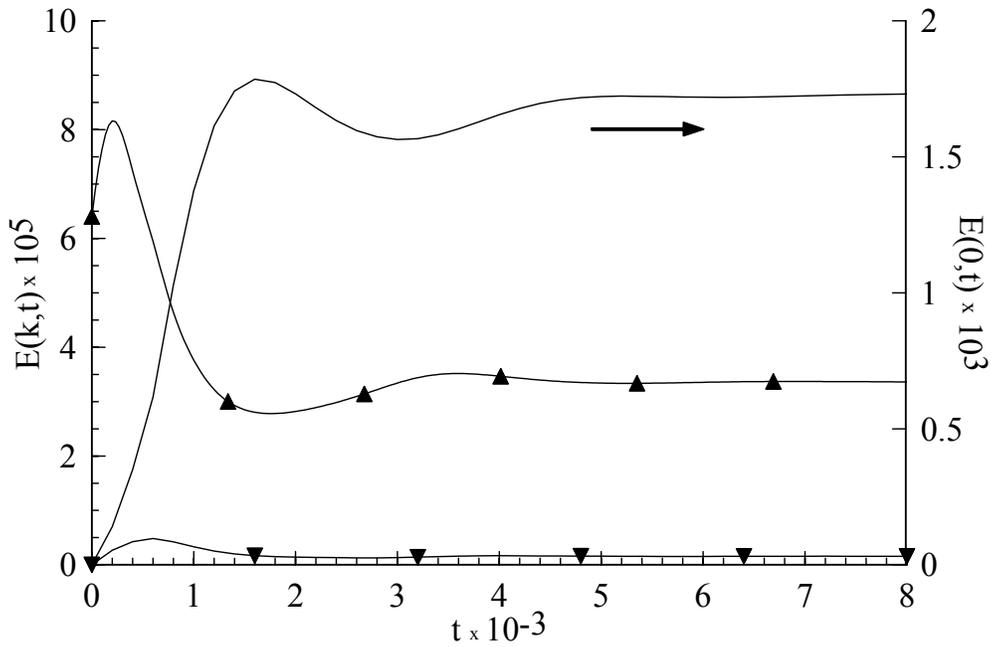

Figure 3b. Evolution of the kinetic energy spectrum for a single initial mode with wavenumber k=0.5.

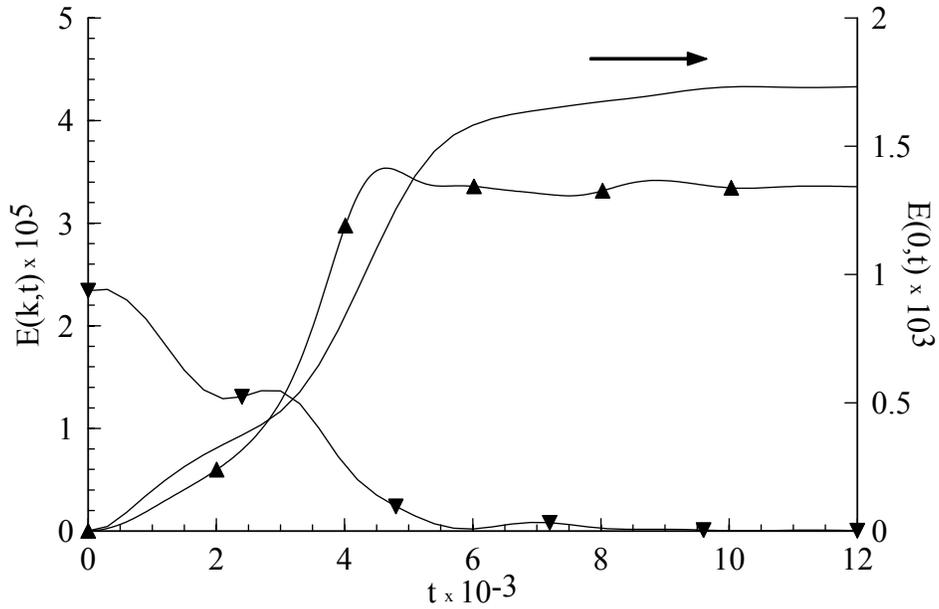

Figure 3c. Evolution of the kinetic energy spectrum for a single initial mode with wavenumber k=0.25.
——— k=0   ▼ k=0.25   ▲ k=0.5

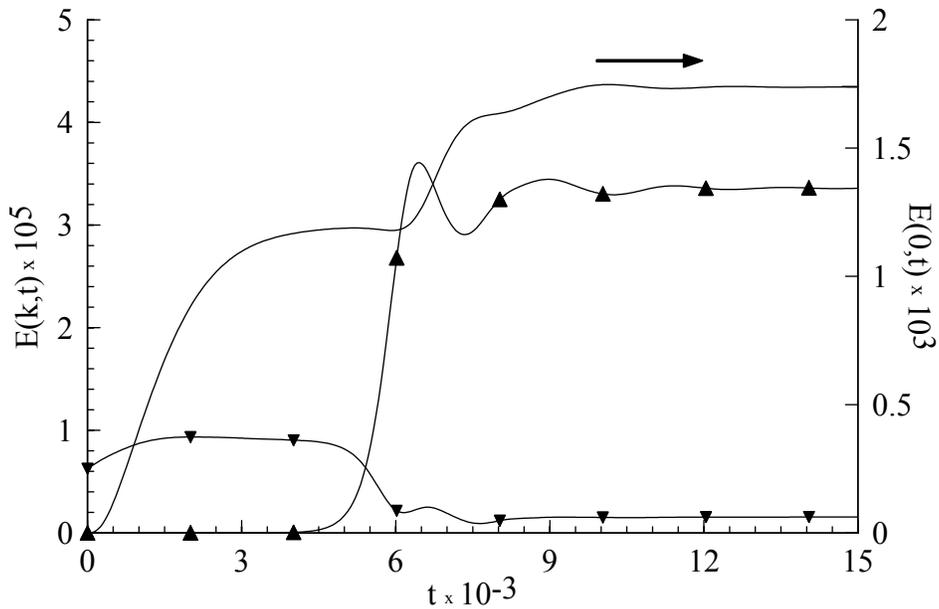

Figure 3d. Evolution of the kinetic energy spectrum for a single initial mode with wavenumber k=1.
——— k=0   ▲ k=0.5   ▼ k=1

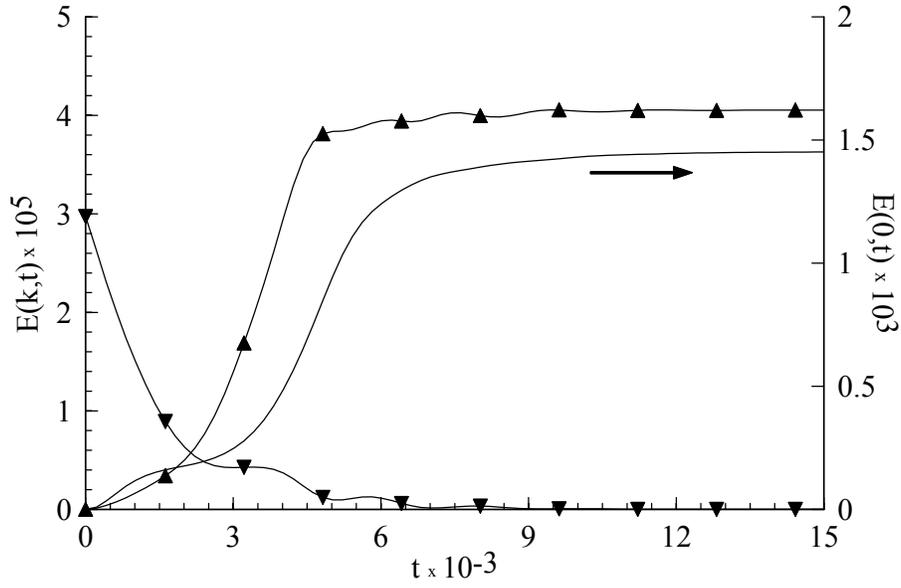

Figure 3e. Evolution of the kinetic energy spectrum for a single initial mode with wavenumber k=0.2.
▼ k=0.2  ▲ k=0.4  —— k=0

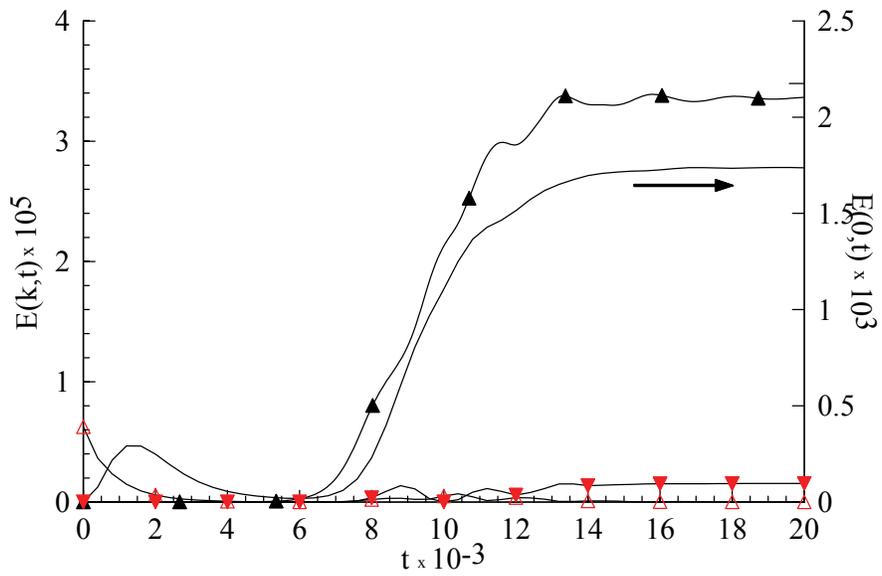

Figure 3f. Evolution of the kinetic energy spectrum for a single initial mode with wavenumber k=1.25.
—— k=0  ▲ k=0.5  ▼ k=1  △ k=1.25

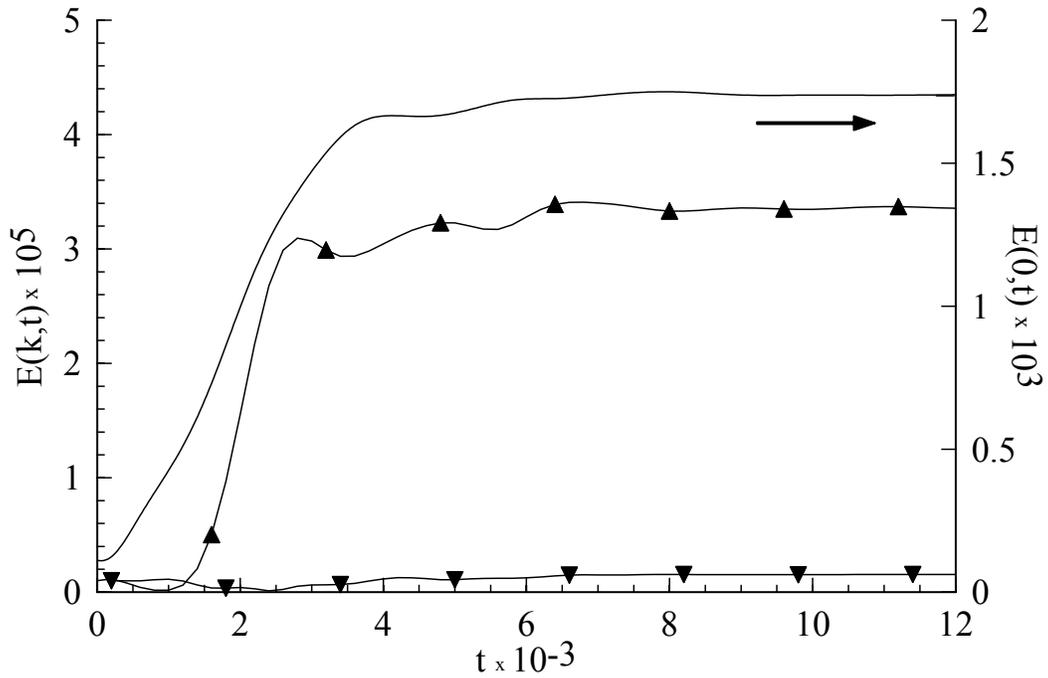

Figure 3g. Evolution of the kinetic energy spectrum for uniform initial condition with amplitude 0.001.
—— k = 0   —▲— k = 0.5   —▼— k = 1

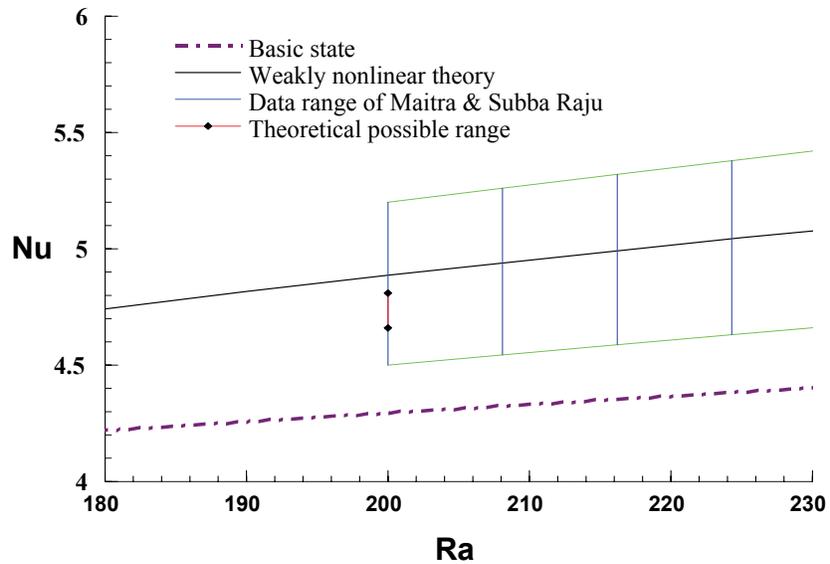

Figure 4. Nusselt number vs. Ra.